# A prototypical Skin Cancer Information System

Ahmed Alharthi
Computer Science Department
Umm Al Qura University
Mecca, Saudi Arabia
Email: adharthi@uqu.edu.sa

Dr. Peter Busch
Department of Computing
Macquarie University
Sydney, Australia
Email: peter.busch@mq.edu.au

Dr. Stephen Smith
Department of Computing
Macquarie University
Sydney, Australia
Email: stephen.smith@mq.edu.au

**Abstract**

*Skin cancer is a common problem in Australia and indeed around the world. Within the domain of e-Health, there appears to be no satisfactory clinical software that follows the flow of a normal skin cancer examination. This paper introduces a system that was specifically designed, coded and implemented to store patient health records as a means of registering the diagnoses of skin cancer along with the treatment. The information system was intended to be web-based, and connect to remote database servers. The implemented system was designed to incorporate features such as inserting procedural details, generating forms and reports with interactive interfaces, yet be relatively unsophisticated to use. We expect the system once fully implemented and on line, will aid in Australia's e-Health industry, delivering more accurate information to doctors and patients in an effort to combat issues involving skin cancer. Other parameters discussed are the need for data encryption of medical records and the role such a system can play in medical information.*

## INTRODUCTION

Electronic Health, eHealth or e-Health is a relatively recent addition to communication technology and information systems in the health sector. The Australian government has spent a substantial amount of money to improve health services through using advanced technology. The Australian Government by way of the Department of Health and Ageing has reformed the health sector and established the Electronic Health Record (EHR) infrastructure as representing a most significant platform (Bartlett *et al.* 2008). The EHR has numerous advantages; it enhances health care and provides a high quality of service; for instance every kind of e-health application is supported by the EHR which shares connections between health care providers, public and private hospitals, the pharmacy, and medical and non-medical specialist providers (Bartlett *et al.* 2008). The EHR can share data in time and provide accurate information, however little development has taken place with regard to systems that handle skin cancer diagnoses specifically. Although healthcare in general is a huge topic area globally (Kuntalp and Akar, 2004), recent research has shown that skin cancer more specifically is an increasing problem in Australia and around the world and is the most common type of diagnosed cancer (Public Health Association of Australia 2009). Doctors require a system to follow the flow of a normal skin cancer examination in order not to miss any diagnoses while the system should reduce the time and cost of repeat examinations and diagnoses, thus increasing the efficiency of e-Health services.

The development a web-based system allows a patients' medical history of skin cancer to be recorded and lesion procedures to be followed through, in order to reduce medical errors and avoid significant misdiagnoses. One doctor articulated this thought in the following words: "we are already running a General Practice (GP) EHR called *Best Practice*, but it does not follow procedural-based medical-practice with accurate follow-up records for each lesion, which is probably the most important aspect of our work in skin cancer." [1] Ideally the system should have efficient interfaces that reduce the need for doctors to type; e.g. a chosen list for data entry should be present negating the need for typing. Moreover, as a part of managing the system, it should have different authorisations for varying roles such as doctors, nurses and receptionists. It is required that the system would; 1) provide a clinical model to highlight the spot or lesion on a patient's body; 2) generate pathology forms and reports from patients' databases; 3) generate

---

[1] Personal communication with Dr. Ian Katz – Southern Sun Skin Cancer Clinic, Hornsby - 1/2/2011.

bills - depending on the part of the body and size of lesion - along with the number of spots or lesions; and 4) upload pathology reports or images, and then record them in the patient's record.

## BACKGROUND: ELECTRONIC HEALTH RECORDS

We may define an EHR as being "a repository of information regarding the health status of a subject of care in computer process-able form, stored and transmitted securely, and accessible by multiple authorised users. It has a standardised or commonly agreed logical information model which is independent of EHR systems. Its primary purpose is the support of continuing, efficient and quality integrated health care and it contains information which is retrospective, concurrent, and prospective" (ISO/TR 20514 p. 2005; Reinecke, 2005). Electronic Health Records have assisted NSW Health in providing a satisfactory level of information to manage and improve the health sector, although some hospitals reject the system because of certain semantic data integration issues, although these issues are currently being addressed.

**Personal Health Record**

The Personally Controlled Electronic Health Record (PCEHR) System which "is a key element of the national health reform agenda around making the health system more agile and sustainable" (NEHTA 2012), was introduced in June 2012. We divide Electronic Health Records (NEHTA 2012) into Personal Health Records (PHR) and Electronic Medical Records (EMR). A PHR is a system partly controlled by individuals in order to manage their own medical records where after a doctor may assist them in meeting their desired goals. As a result, the PHR can assist physicians to make more informed decisions whilst improve the sharing of health records (Tang *et al.* 2006). If however individuals keep their records up-to-date but make mistakes by entering inaccurate data, there will be consequences in their treatment methodology (Liu *et al.* 2011). In the U.S. literature many examples are to be found of the benefit of PHRs, from being able to receive messages from different health care providers to seeing results of lab tests or making appointments, to exchanging information with providers. Another example of PHR benefits are noted by Caligtan and Dykes (2011) with regard to the Dana Farber Cancer Institute at the Brigham and Women's Hospital as well as at the Massachusetts General Hospital where a PHR gateway is provided to manage chronic diseases. It is worth noting that in the U.S. there are currently over 65,000 PHR record holders users, and by 2014 it is estimated that most Americans will use EHRs that incorporate PHRs.

Although the EHR system has numerous beneficial features, it does have some constraints. One study by Liu *et al.* (2011) assessed three websites: Google health, Microsoft HealthVault (Do et al., 2011), and WorldMedCard and found issues with regard to: *usability and complexity* - systems often have complex features, with a number of functions and tasks that led users to believe they were wasting rather than saving time; *terminology* - users often tended to have low health literacy, which resulted in potential for entering inaccurate data; *familiarity and comfort* - when a system is different or unfamiliar with other products, it may affect the use of the system; *communication and integration* - where users were concerned with the ability to share data and integration; *privacy, security, and trust* – where trustworthiness in the system is diminished; and finally tying in with *terminology* – issues with regard to that of *ensuring accurate data*.

**Electronic Medical Record**

An Electronic Medical Record (EMR) interacts with Hospital Information Systems (HIS), Radiological Information Systems (RIS) and Picture Archiving and Communication Systems (PACS) to collect patient data in one place. When clinicians diagnose patients, they typically require x-rays; consequently patient results will be stored in the EMR which in turn will document and provide evidence for clinicians to choose appropriate treatment (Ebadollahi *et al.* 2006). Data in EMRs can be categorized as text, image, video and other media formats such as signals of electro-encephalograms. Other benefits relate to medical students being able to enhance their education through multimedia data and observing how qualified doctors treat incurable diseases, providing the students with a greater understanding (Ebadollahi *et al.* 2006). Data from EMRs can be captured and stored from various wards, departments and medical devices throughout the hospital infrastructure as well as relevant information systems. Typically patients can access their records via the internet from any location and can print lab results through access via a portal, using a user-name and password.

Criticisms also exist of EMRs; Patrick (2009) discusses the effectiveness and impact of Cerner FirstNet in NSW hospitals, where some physicians rejected using the system because it was very slow and required a lengthy period of time to respond. Furthermore physicians felt they had extra work to do in typing and thought they were wasting time instead of saving time. Some individuals refused to use EMRs because they were also obliged to pay for use of the system (Tang *et al.* 2006). Ebadollahi *et al.* (2006) presented three main challenges for EHRs; the first challenge was the *unifying architecture*. When EHRs are integrated with other systems such as PHRs and other institutions and providers' systems, there may be *semantic integration* issues. Different interfaces represent a case in point. In addition, the *structure of information* represents another difficulty because there are various data sources and methods to capture data from different technologies, for example different databases often need to be mapped. EMR systems also tend to lack effective *analysis engines* – other than the traditional diagnosis abilities of medical staff.

## SOLUTIONS

Patrick (2009) suggested ways of increasing the effectiveness of EHRs. First, familiar interfaces need to be designed, as well as recognition for reduced typing in order to decrease the possibility of errors. Secondly open source systems are considered desirable. Tang *et al.* (2006) further noted education could solve common problems in EHRs - for when individuals acquire the knowledge to use EHRs as early as their school years, they will be better positioned to manage their records and improve their own health literacy. For example young users may want to know the difference between diabetes type 1 and type 2; they might want to know how to record their continuity of medicine taken over time, or their nutrition with regard to food quantity and quality; while more senior level students in medical school should know how to use the EHR system as part of their studies; e.g. when one doctor who studied a EHR writes reports or enters accurate data, other doctors can benefit from this patients' record and take considered action. Liu *et al.* (2011) also suggest some solutions to current problems: categorization of menus will lead to reduced confusion; certain descriptions of terminology might prove helpful; if users customize their own records, it would make them more usable; and designers could improve the accuracy of the data. With regard to security issues specifically, Yee and Trockman (2006) note EHRs data can be encrypted while being exchanged through a system and suggest approaches to managing access for multiple users via different keys that implement varying access restrictions. One solution is a patient-held electronic health-record wallet-card. When patients approach health-care providers, they can give them the card and have it updated with further information. If a patient wishes to see their information online, they should have a username and password to access their account; the card could be used in emergency situations as well.

## SCIS: SYSTEM OVERVIEW

Taking the above e-Health parameters into account, this paper discusses the implementation of a web-based Skin Cancer Information System (SCIS) incorporating EMRs. Users require access from any device having browsers and connecting to the Internet. In addition the SCIS should generate reports, bills and forms and connect with other records such as PHR, therefore the SCIS will include shard EHR specifications, clinical information, clinical terminologies, user authentication, standards implementation, and longitudinal health records. The Australian Government is heavily promoting e-Health and has standardised the system interoperability for EHRs and in turn the SCIS will be comprise a part of e-Health in Australia. The system should include: *patent information*: including general details such as full name, address and contact details; *patient history* including health information such as previous health details, allergies and medication used; *management* that is to say observing the flow of health information including employees printing reports; *reception* or adding patients' details and organizing waiting lists; *clinical models* of the human body illustrating the position of lesions on patients; *staff information s*uch as contact details; and finally the ability to *generate reports, forms and bills*.

**Approach**

The methodology adopted a design or service-science approach (Kaner and Karni, 2007) using a combination of waterfall and iterative development which took the output of a Software Requirements Specification (SRS – table 1), elicited from the client doctor. Page length limits prevent detailed discussion of the approach adopted, but the requirements were implemented by designing, coding and testing iteratively (Sommerville, 2010), continually engaging with the doctor. Requirements elicited were the following:

Table 1: Functional requirements of the SCIS including risk, effort and priority

| REQ # | Description | Risk[1] | Effort[2] | Priority[3] |
|---|---|---|---|---|
| R1 | All users *Login* - shall provide system access having suitable security services. This access will have various levels that depend on user authorization. | H | M | C |
| R2 | Physician and Nurses *Create New Record* - shall provide physicians and nurses with the ability to create a new record for patients for the first time | M | M | I |
| R3 | Physician and Nurses *Create New Problems* – shall provide physicians and nurses the ability to create a problem in a patients' record. When patients have a problem, the problem will be described and diagnosed. | M | M | I |
| R4 | Physician and Nurses *Create Visit* – shall enable physicians and nurses to record each visit that may have various problems and different procedures. | L | M | I |
| R5 | Physician and Nurses *Edit Record* – shall enable physicians and nurses to edit records by updating or adding more information. | L | M | I |
| R6 | Physician and Nurses *Insert Procedure* – shall enable physicians and nurses to select appropriate procedures for one problem or more than one. | M | H | C |
| R7 | Physician and Nurses *Finalize Procedure* – shall enable physicians and nurses to complete record and finalize the procedure. | M | L | U |
| R8 | Physician and Nurses *Access Patients' Record* – shall enable physicians and nurses to view record and previous problems with their procedures and any previous history that was recorded. | H | H | C |
| R9 | Physician and Nurses *Allocate pathology report to procedure* – shall enable physicians and nurses to allocate any pathology report to its procedure in a patients' record. | M | L | I |
| R10 | Physician and Nurses *Upload documents and image* – shall enable physicians and nurses to upload documents and images to a patients' record. | L | M | I |
| R11 | Physician and Nurses *Generate and Print form* – shall enable physicians and nurses to generate forms such as, taking a test and printing it. | M | M | I |
| R12 | Physician and Nurses *Generate bill* – shall enable physicians and nurses to generate bills and print them. | L | M | I |
| R13 | Physician and Nurses *Hold or Un-hold bill* – shall enable physicians and nurses to hold bills until the result appear, then un-hold them to continue the process. | L | M | I |
| R14 | Physician, Nurses and Receptionist *Print bill* – shall enable physicians, nurses and receptionist to print bills. | L | L | U |
| R15 | Physician, Nurses and Receptionist *Create patients' information* – shall enable physicians, nurses and receptionist to create patients' information. | H | M | C |
| R16 | Physician, Nurses and Receptionist *Edit patients' details* – shall enable physicians, nurses and receptionist to update patients' information. | M | L | I |
| R17 | Receptionist *Create waiting list* – shall enable receptionists to create waiting lists and update them. | L | L | U |
| R18 | All users *Search* – shall enable all users who have authorisation to look at different information via a search feature, including patient and staff information. | H | M | C |
| R19 | Administrator and Manager *Print report* – shall enable administrators and managers to print various reports. | L | M | I |
| R20 | Administrator and Manager *Create new staff account* – shall enable administrators and managers to create new staff account and enter their details. | M | L | I |
| R21 | Administrator and Manager *Edit staff's details* – shall enable administrators and managers to update staff details. | L | L | U |
| R22 | Administrator *Manage role* – shall enable administrators to locate staff authorization. | H | L | I |
| R23 | Administrator *Create Centre's information* – shall enable administrators to establish the centre's information and entering important details such as connecting details. | L | L | U |

1. *Risk*: High (H)/Medium (M)/Low (L) 2. *Effort*: High (H)/Medium (M)/Low (L) 3. *Priority*: Critical (C)/Important (I)/Useful (U)

## SYSTEM DESIGN

The SCIS is web-based, hence the system is hosted by using a web-server, but the client should not need to worry about security and backup (Jazayeri, 2007; Linthicum, 2004). Apache Web Services, a Tomcat Application Server and MySQL Database will be used for the SCIS because they are open source, and provide a high level of community support through seamless development initiatives. To satisfy the client's demands and to conduct business at light-speed, the client can use the internet to connect to remote database servers using a Web Server (figure 1); this web-server will deal with the information and exchange between applications and the system. As may be seen in figure 2, the web-server based approach was recommended as a standard for electronic health records by the National e-Health Transition Authority (NEHTA) Australia (2006) because of the support for appropriate protocols.

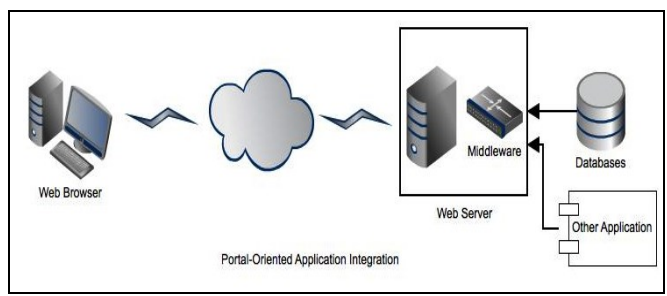

Figure 1: Portal-Oriented Application (source: Linthicum, 2004 p. 20)

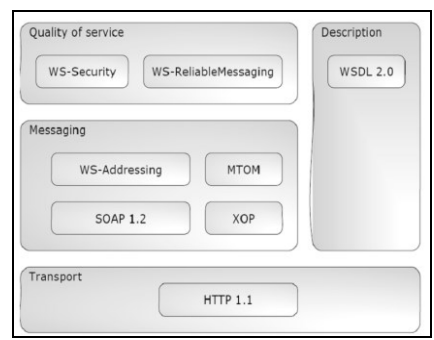

Figure 2: Recommended Web Services Standards (source: NEHTA, 2006 p. 5)

**User Interface**

The first main interface (figure 3) illustrates the patient's clinical and procedural details as well as headers providing navigation to cross inside the clinical portal. For example, there should be waiting lists, the ability to edit patient's details and insert the patient's general medical information. Figure 4 illustrates the administrative interface providing a menu on the left to guide between pages such as inserting or editing patient or staff details. The third main interface (figure 5) provides interfaces such as a log-in page. Finally the headers in the first and second interface provide the details on who is accessing the system, with three icons to navigate between the physician, the manager and receptionist's portals.

**The database**

The client provided permission to access their server, create a subdomain and install the database that was designed in MySQL (jQuery Project, 2012). The system was tested with certain transactions, for example the insertion of staff details.

**The management interface**

The management and reception components were designed using *Muse* web designer software. An example of the login page is provided (figure 5 below). Database connection code using PHP between the client and the server side is provided in figure 6. After HTML files were tested via unit testing in local storage, they were uploaded and tested with PHP files and the database.

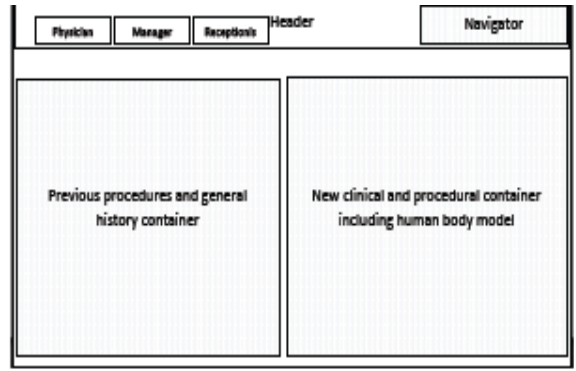

Figure 3: The clinical and procedural interface of the SCIS

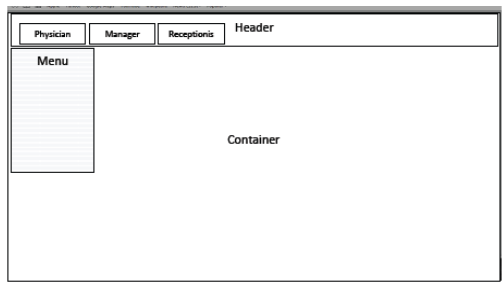

Figure 4: The administration and receptionist interface of the SCIS

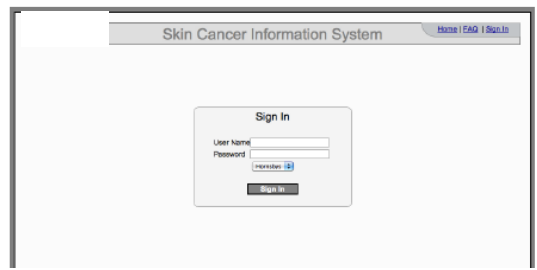

Figure 5: The log-in interface of the SCIS

**Patient and user forms**

Designing and coding other pages continued in the third stage. Examples of work in this space include inserting and storing new staff and patient details (figure 7). The data was tested after inserting into the database - when bugs were found, they were fixed and tested again. The client was involved at regular intervals in the development of the system.

**The clinical portal**

In last stage (figure 7), the clinical portal which included the human body illustration as well as all patient health details (figures 8 and 9) were designed and developed with HTML, PHP and JavaScript. Also jQuery as a library in JavaScript was used to hide or show certain parts of clinical and procedural information; for example, when users fill in clinical details - they can click on the <<next>> button which includes a function to hide clinical information, but show procedural details.

Figure 6: The configured connection PHP file of SCIS

Figure 7: Inserting a new account page for staff in the SCIS

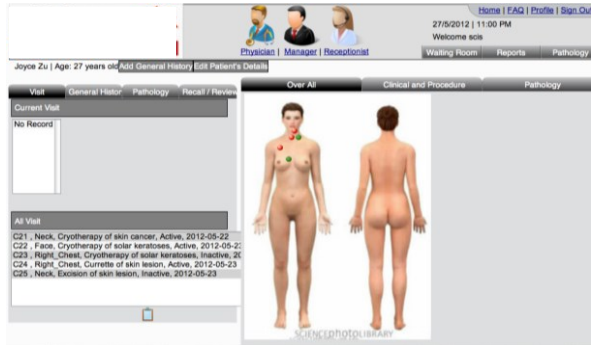
Figure 8: The clinical portal of SCIS

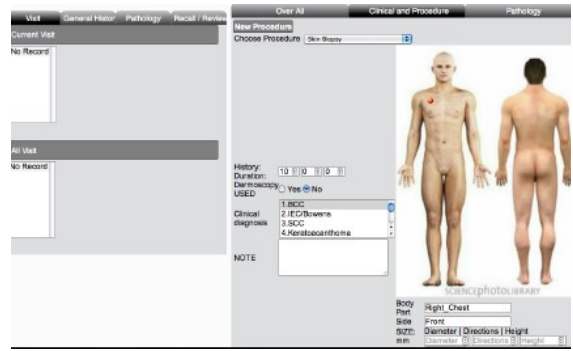
Figure 9: Choosing a new procedure

**Verification and Validation**

Various methods were applied to ensure data verifiability and validity. First of all, HTML5 forms were implemented for input fields where the user inserts information. For example, email and date input fields used Email data-type and a date-type for the date of birth or other date input (figure 10). Although some browsers do not support it, it was recommended to use Google Chrome 19.0 and Safari 5.1. Secondly, to insure the process data was correct, drop-down-lists, radio buttons and checkboxes were also used.

**Security Implementation**

To apply security to the SCIS, it was requested the user log-in to the system via a user-name and password which was encrypted by applying MD5 (a Message-Digest function in PHP). When the user logs in, their username and password will be sent to a webserver which encrypts the password before comparing it with what is in the database. Moreover there are three levels of security to use the system, as each group of users have specific pages. For instance, receptionists cannot access clinical portals, but physicians can access receptionist portals. Finally images are stored in the database instead of being saved as files in the server. In this way no-one can see images without authorisation to the database. Finally, when the client backs up the database, images will be saved securely and efficiently (Wong and Huang, 1996).

# TESTING

After implementing the code it was tested; if it passed then it was tested with a mistaken case in order to ensure validation and verification.  The codes was then integrated with the system and again verified through correct and incorrect cases. Figure 11 illustrates the concept of unit and integration testing applied to the SCIS which was conducted iteratively through the four major stages of implementation. The former approach is a white-box test while the latter method is applied as a black-box test. Finally, after testing the SCIS and cleaning the test data, the SCIS was deployed.

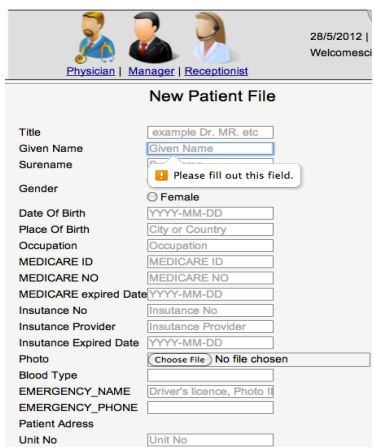
Figure 10: The clinical portal of SCIS

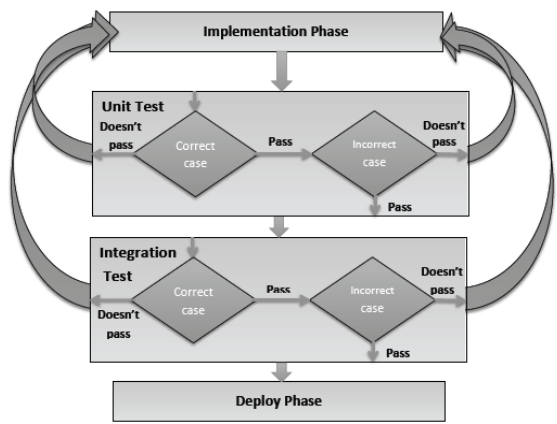
Figure 11: The test methods of the SCIS

One of the main contributions of the system development was the ability to translate the system user's functional requirements and prioritise them in the system's development. Examining figure 12 the reader can see that R1 (recall table 1 earlier): suitable security services; R6: the ability of doctors and nurses to be able to insert an appropriate procedure into the SCIS; R8: the ability of medical staff to view previous patient histories; R15 the ability to enter new patient information; and R18: the need for the system to have a search feature – all show the highest level of requirement in the system design. Other requirements (table 1) feature less prominently in our figure 12. This follows-on in that the medical staff priorities for patient care naturally became the highest priority in the system development while lower risk, effort and priorities requirements were ranked down the list. Certainly the link between medical imperatives, patient history and clinical care was a key outcome of this project. Initial feedback elicited from medical staff using the system indicate the requirements given above were met, although the system is currently prototypical.

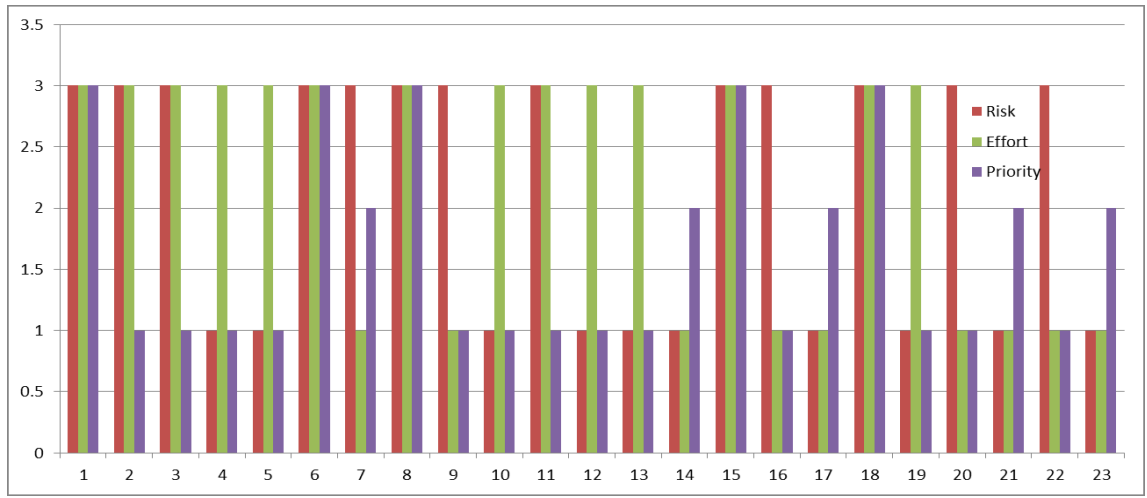

Figure 12: Functional requirements risk and effort profile
(X axis = requirements given in table 1; Y axis: 1= risk, 2=effort, 3 = priority)

## DISCUSSION

Our results confirm that users (physicians) should be involved during the development of EHRs because while we are developing the SCIS, we faced some difficulties in understanding medical terminology as part of the design process. For example, a doctor upon seeing the clinical model, asked for the insertion of different colours relating to a lesion's condition, to reduce the time in assessing a patient as noted previously by Saugeon, Guillod and Thiran (2003). In addition, the doctor requested the flow of the system be accurate because it would obviously affect a patient's life. When doctors create a new shave biopsy; the flow should be inserting clinical details followed by a patient's consent and then procedural information should be included followed by a final confirmation notice. Clearly the doctor as domain expert was a necessary partner in the design process.

Interactive interfaces were designed that resulted in reducing typing errors and increased data accuracy. When users created a new procedure, they chose a procedure from a drop-down list. They did not need to type procedural names, as typing a procedural name might lead to confusion concerning a procedure as well as taking more time such as in type *cryotherapy of Skin Cancer*. Therefore, interactive interfaces would decrease mistyping and avoid imprecise data, tying in with Ebadollahi *et al.*'s (2006) concerns of a unifying architecture and Patrick's (2009) ways for increasing the effectiveness of such a system.

Studies show data should be captured and stored and categorized as simple text, image, video and other media formats, but we strongly suggest processing these as multimedia data. For example, in skin cancer prognosis, doctors take different photos of certain spots from various sessions; these photos need to be processed through image processing to find the differentiation between images, which in turn can show the progress of medication. Naturally such a process aids doctors in decision making.

In line with Yee and Trockman's (2006) concerns over the encryption of medical records, we also tested the security level and encryption, observing that various privileges could be applied in EHRs. Doctors can have a certain level of security to medical data, and managers also can have another level to access data without seeing specific medical details. For example, managers may generate reports containing a number of procedures for patients, without being aware of procedural details. In addition data and images were encrypted; this assisted in decreasing any threats that may affect data, because if data and images are not encrypted, they may be attacked and easily changed by hackers, as noted by Liu *et al.* (2011).

Last but not least, the doctor and other personnel involved in providing feedback on the system were favourable to it being used for educational purposes at the local hospital, where the educational advantages of e-Health systems were noted by Tang *et al.* (2006).

## LIMITATIONS AND FUTURE WORK

As with all information systems there are aspects that can be improved over time. To date, limitations to the system including a recognition the design of the system could be improved through the use of an iOS platform (a mobile operating system developed by Apple), and also integrating the system with other systems. Other limitations were that the system has to be hosted in a secure place or at least locally in Australia at this stage, in order to keep patient records secure. Another limitation is that all procedural details are not encrypted. The interfaces currently are not ideal, more professional Cascading Style Sheets (CSS) are required, such as applying Class and ID tags for each element in the HTML files, although some have an ID or Class tag. The database does not incorporate triggers to verify data before storing such data; for instance the database ideally requires a trigger to capitalize the first letter of given name and last name; as well as another trigger to ensure there is no duplicated data, for if the patient has checked-out a file while the receptionist adds a new file by mistake, the user will not realize this and the system currently will accept the change. The system also does not include pathology and billing components; the SCIS should have a pathology 'inbox' to receive results from the laboratory. The system could also be improved through sending SMSs or emails for prescriptions and reminders and so forth.

Finally, the project should not standalone; rather the SCIS should be integrated with electronic health records and e-government initiatives as a whole. More work is required on the system security as well as solving the difficulty in standardising electronic health records. There are also usability issues requiring further studies with regard to doctors and nurses refusing to use EHRs.

## CONCLUSION

Many governments around the world have further developed their health sectors through implementing Electronic Health Records (EHR) which are an application for storing and transmitting health information securely. They can integrate with various subsystems; for instance one of these subsystems was that of Personal Health Records (PHR) that are controlled by individuals in order to enter and update their information or to see their health record and reports. Another part of HER comprise Electronic Medical Records (EMR), which are systems existing in hospital or by health care providers. Additionally, EHRs possess some handicaps relating to usability, privacy, security, and trust. Some constraints facing EHRs could be solved through education via public schools, medical schools, universities and through investigation and through research. This paper introduced a Skin Cancer Information System (SCIS) for tracking skin cancer lesions and the procedures applied to manage them. The introduction of such a system is innovative as few such examples exist globally and in the current literature. The client required a system having functional and non-functional requirements and being web based. The web-based application connects to a remote database utilizing a web server. Having discussed the need for such a system and the design of it, we conclude by saying the design process needed to involve physicians in the design of interactive interfaces. We also noted the importance of using multimedia rather than simple data types given the content of the system, as well as applying various levels of authorisation given the nature of medical records requiring encrypted data.

## ACKNOWLEDGEMENTS

We wish to acknowledge the contribution of Dr. Ian Katz, Southern Sun Skin Cancer Clinic, Hornsby N.S.W.

## COPYRIGHT